\documentstyle[preprint,aps]{revtex} 
\begin{document}
\preprint{UM-P-95/113, RCHEP-95/25} 
\draft 
\title{POSSIBLE ELECTRIC CHARGE
NONCONSERVATION AND DEQUANTIZATION IN $SU(2) \times U(1)$
MODELS WITH HARD SYMMETRY BREAKING}
\author{A.Yu.Ignatiev\cite{byline1} and
G.C.Joshi\cite{byline2}} \address{Research Centre for High
Energy Physics, School of Physics, University of Melbourne,
Parkville, 3052, Victoria, Australia} 
\date{To appear in {\em Physics Letters B}} 
\maketitle
\begin{abstract}

 We study a novel type of extensions of the Standard Model
which include a hard mass term for the U(1) gauge field and,
optionally, the additional scalar multiplets spontaneously
violating the electric charge conservation.  Contrary to the case
of abelian massive electrodynamics, in these theories the
massiveness of photon necessarily implies non-conservation (and
also dequantization) of the electric charge (even in the absence
of spontaneous breakdown of the electromagnetic symmetry).  On
the other hand, unexpectedly, there exist models with charge
non-conservation  where it is possible to keep the photon mass
zero (at least, at the tree level).

\end{abstract} 

1. In the past, there have been many papers exploring the possibility
that the photon may have non-zero mass \cite{GN}. At first, 
these works were
made in the context of massive electrodynamics, which is an abelian
U(1) theory with the added photon mass term 
${1 \over 2}m^2 A^2_{\mu}$. The characteristic feature of such
 a theory is that the conservation of the electric charge  is not
violated by the presence of the photon mass term. The reason is
that the photon mass term violates the local gauge invariance but
not the global one. Thus, massive electrodynamics suggests that it 
it is possible to have a massive photon along with the exact 
conservation of the electric charge.

Later, there emerged another class of theories with massive photon:
non-abelian gauge theories with {\em electric charge
 non-conservation} \cite{we1,okun,we2,s,ts,Qnoncons,mtt,mn}.
 The primary emphasis of these works was not on
the massiveness of photon , but on the study of the possible electric
charge non-conservation in gauge theories. Yet the massiveness of 
photon appeared to be an automatic consequence of the violation
of the electric charge conservation. One of the important discoveries 
made in those works was the close
relation between the two ideas: electric charge (non)conservation and
electric charge (de)quantization \cite{we1,we2,Mel}.

Given all previous works, one important question still remains
unexplored: is it possible to have massive photon {\em and exact 
electric charge conservation} in realistic theories? Of course, it
is possible within U(1) massive electrodynamics, but it is {\em not} 
a realistic theory. What we are interested in is this: can the
 Standard $SU(2) \times U(1)$ model be extended or modified in such
a way as to have both the massive photon and the exact electric
charge conservation simultaneously.

At first sight all we have to do is to give a hard mass to the U(1)
gauge boson $B_{\mu}$. That would presumably make photon massive
without spoiling the electric charge conservation. However, we will
show that it turns out {\em not} to be the case.

A related but different question that we are going to consider is
this: can we have {\em a massless photon} in a theory with electric
charge non-conservation? One reason to ask this question is the
very stringent experimental bound on the photon mass: $m_{\gamma}
< 10^{-24}$ GeV or even $10^{-36}$ GeV \cite{partdata}. 
This bound places a very
tight constraints on any theory with electric charge 
non-conservation and one is naturally curious how to evade it.

Naively, one might think that the answer to the 
above question is negative. Yet in
this work we will construct examples of realistic $SU(2) \times
U(1)$ models in which the photon mass is {\em zero} at the tree
level but the electric charge is {\em not} conserved\footnote{Note
that the elaboration of these models beyond the tree level lies
outside the scope of this paper. Thus at this stage we cannot
rule out the possibility that the ultimate answer to the above 
question should be negative}.

2.Let us first consider a model which differs from the Standard Model
 only in
one point: its lagrangian contains a mass term for the $U(1)$ gauge
field $B$ (before spontaneous symmetry breaking):
\begin{equation}
{\cal L}'={\cal L}_0 +{1 \over 2}m^2 B^2_{\mu},
\end{equation}
where ${\cal L}_0$ is the Standard Model lagrangian.

After spontaneous symmetry breaking, we diagonalize the gauge boson
mass matrix and obtain the physical fields A and Z which we 
identify
with the photon and Z-boson:
\begin{eqnarray}
A^3_{\mu} &=& Z_{\mu} \cos \theta ' + A_{\mu} \sin \theta ' \\
B_{\mu} &=& A_{\mu} \cos \theta '  - Z_{\mu} \sin \theta '.
\end{eqnarray}
where the mixing angle $\sin^2 \theta'$ is different from
the Weinberg angle of the Standard Model ($\sin^2 \theta $):
\begin{equation}
\sin^2\theta ' =\sin^2 \theta + {m^2 \over M_Z^2} (1 - {e^2 \over 
\sin^2
\theta} + e^2 - \sin^4 \theta) \approx \sin^2 \theta + 
0.64  {m^2 \over M_Z^2}.
\end{equation}

Note that the photon acquires a non-zero mass:
$M_{\gamma}=gm$.
Having obtained the mixing angle we can now write down the
electromagnetic part of the lagrangian  ${\cal L}^{'}$:
\begin{eqnarray}
{\cal L}^{'em} &=& {\cal L}_l^{'em} + {\cal L}_q^{'em} \\
{\cal L}_l^{'em} &=& A_{\mu}[{1 \over 2}(g\sin\theta' - 
 g'\cos\theta'
)\bar{\nu}_{L}\gamma^{\mu}{\nu}_L
-{1 \over 2}(g\sin\theta' + g'\cos\theta' )
\bar{e}_{L}\gamma^{\mu}{e}_L \nonumber\\
&&- g'\cos\theta' \bar{e}_{R}\gamma^{\mu}{e}_R] \label{35} \\
{\cal L}_q^{'em} &=& A_{\mu}[{1 \over 2}
g\sin\theta'(\bar{u}_{L}\gamma^{\mu}{u}_L -
\bar{d}_{L}\gamma^{\mu}{d}_L) + g'\cos\theta' ({1 \over
6}\bar{u}_{L}\gamma^{\mu}{u}_L +{1 \over
6}\bar{d}_{L}\gamma^{\mu}{d}_L + \nonumber\\
&& {2 \over
3}\bar{u}_{R}\gamma^{\mu}{u}_R - {1 \over
3}\bar{d}_{R}\gamma^{\mu}{d}_R)], \label{36}
\end{eqnarray}
where $g$ and $g'$ are SU(2) and U(1) coupling constants
(the rest of the notation being self-explanatory).
Based on this formula, we can arrive at an important conclusion: as
soon as the equality $g \sin \theta = g' \cos \theta$  is broken, the
electromagnetic current conservation is violated immediately.
 To avoid confusion, one essential point needs to be emphasized here.
We have defined the electromagnetic current (and thereby the electric
charge) as the current interacting with (i.e. standing in front of)
the electromagnetic field $A_{\mu}$. Naturally, one can ask about the
standard fermion electromagnetic current of the form
\begin{equation}
j_{\mu}= e(-\bar{e} \gamma_{\mu}e + {2 \over 3} \bar{u} \gamma_{\mu}u
- {1 \over 3} \bar{d} \gamma_{\mu}d). \label{37}
\end{equation}

 Although this current is still conserved in the present model
, it  unfortunately becomes devoid of physical meaning,
because all physical processes and experiments are based on the
interaction between the charges and electromagnetic fields; therefore
in the framework of the present model we have to attach physical
meaning and reserve the name "electromagnetic current" for the 
current
of Eq.(\ref{35}) and (\ref{36}), rather than that of Eq. (\ref{37})

To summarise, this theory features three fundamental deviations from
the Standard Model: massiveness of photon, the electric charge
dequantization, and the electric charge non-conservation.

Considering the experimental limits on the parameter $m$
we note that the experimental upper bound on the photon mass gives, 
by
far, the strongest constraint on the value of $m$. It has been
established that the photon mass should be less than $ 10^{-24} $  
GeV or even
$ 10^{-36} $ \cite{partdata}. Therefore, we 
find that 
the parameter $m$ cannot
exceed $2 \times 10^{-24}$ GeV or $2 \times 10^{-36}$ GeV.   
With such small values of the parameter $m$,
the charge dequantization and charge non-conservation effects are
expected to be too small to be observed. 
(for a detailed discussion of experimental constraints on models
with electric charge dequantization, see 
\cite{bv}).

Note that this model (with no fermions) was first suggested in Ref. 
\cite{clt} under the name
of "hybrid model". The authors of Ref. \cite{clt} were motivated by 
the
systematic search for renormalizable gauge models beyond the standard
$SU(2) \times U(1)$ model. As concerns the renormalizability of the
model which is certainly a very important issue, it has been proved 
in
Ref. \cite{clt}  that the theory posesses the property called tree 
unitarity
which is a weaker property than renormalizability. We are not aware of
any work which would further address the problem of renormalizability
of this type of models (cf. Ref.\cite{delb}).
Although it may appear to be of academical
rather than phenomenological character, this work would certainly be
very desirable because it would include or exclude  a whole new class
of gauge models from the set of renormaliz able gauge theories. (Note
that we do not share the belief that non-renormalizability of a theory
automatically makes it physically uninteresting.)

3.Let us now add to the previous Lagrangian  a piece
containing the scalar singlet field $\phi_1$ with the electric charge
$\epsilon$ (which coincides with the hypercharge in this case):
\begin{equation}
\label{38}
{\cal L}_1 = {\cal L}_0 + {1 \over 2}m^2B^2 + |(\partial_{\mu}  -i {g'
\over 2} B_{\mu}) \epsilon_1 ) \phi_1|^2 + P(\phi_1 , \phi). 
\end{equation}
Now, assume that the field $\phi_1$  has non-zero vacuum expectation
value $v_1$: $\langle \phi_1 \rangle = v_1$. Then, after spontaneous
symmetry breaking 
and performing the diagonalization as before, we obtain the mass 
of the
physical photon to be:
$M^2_{\gamma} = g^2 (m^2 + {1 \over 2} g'^2 v_1^2
 \epsilon_1^2 ) $.
Thus, the formula for the photon mass (squared) consists of 
two
contributions: the first is proportional to $m^2$ (``hard mass'') and
the second is proportional to $v_1^2$ (``soft mass''). Nothing seems
to prevent us from considering {\em negative} value for $m^2$
(without loss of generality, $v_1^2$ can be always made positive
by an appropriate gauge transformation).
Thus we are led to a very interesting possibility: to
choose these two parameters in such a way that they exactly cancel
each other so that the photon remains massless\footnote{Here, we 
disregard a possible appearence of a Nambu-Goldstone boson. One may 
expect that its manifestations would be sufficiently suppressed, but
even if they were not, the model could be modified in analogy with
Ref. \cite{mtt}.}
(at least, at the tree
level):
\begin{equation}
m^2 + {1 \over 2} g^{'2} v_1^2 \epsilon_1^2 = 0. \label{41}
\end{equation}
It can be shown that if the condition (\ref{41}) is satisfied,
the Z-boson mass is not shifted at all.
Now, do we obtain the electric charge non-conservation or
dequantization in the fermion sector, in analogy with the result of
Section 2? Unfortunately, the answer is: no. The reason is this: the
calculation of the Weinberg angle in this model (denoted by
$\theta_1$) shows that this angle is {\em exactly equal} to the
Weinberg angle of the Standard Model:                                 
$\sin^2 \theta_1 = \sin^2 \theta $.
Note that this exact equality has been obtained without assuming $m^2$
or $v_1^2$ to be small (but, of course, assuming that the condition
of photon masslessness, Eq. (\ref{41}) holds.)
From this equality it follows that the fermion electromagnetic current
 in this
model remains exactly the same as in the Standard Model:
$j_{\mu}= e(-\bar{e} \gamma_{\mu}e + {2 \over 3} \bar{u} \gamma_{\mu}u
- {1 \over 3} \bar{d} \gamma_{\mu}d)$.
In other words any effects of the electric charge non-conservation or
dequantization are absent {\em in the fermion sector}. Here we would
like to stress an essential point: the absence of these effects in the
fermion sector {\em does not mean} that they are absent altogether.
One should not forget that giving the vacuum expectation to the
charged scalar field $\phi_1$ leads to the electric charge non-
conservation {\em in the scalar sector}. However, from the
phenomenological point of view, these effects are much harder to
observe. Such effects would be similar to those arising in a model
with charged scalar field but without the $m^2$ term. Models of such
type have been considered in the literature before and we do not
intend to go into details here.

Thus we see that in the context of the
model with the lagrangian (\ref{38}), vanishing of the photon mass
 leads to vanishing effects of
charge non-conservation and charge dequantization {\em in the fermion
sector} (but not {\em in the scalar sector}
).

4.Let us  now change the
singlet into the scalar {\em doublet}, again violating U(1) symmetry;
the rest of the model will be the same. Thus, the lagrangian of our
new model reads:
\begin{equation}
{\cal L}_2 = {\cal L}_0  +  {1 \over 2}m^2B^2 + 
|(\partial_{\mu} -ig {\tau^a \over 2} A^a_{\mu} -i {g'
\over 2} (1 + \epsilon_2)B_{\mu}) \phi_2|^2 + P(\phi_2 , \phi). 
\end{equation}
where the electric charges of the scalar doublet are:

\begin{equation}
 Q(\phi_2) =\left( \begin{array}{c}
1+ {\epsilon_2 \over 2} \\ {\epsilon_2 \over 2}
\end{array} \right). 
\end{equation}

We break the electromagnetic symmetry by assuming
\begin{equation}
\langle \phi_2 \rangle = {1 \over \sqrt{2}} \left( 
\begin{array}{c}
0 \\ v_2
\end{array} \right). 
\end{equation}
After the spontaneous breakdown of symmetry we can find out that
the condition for the photon to be massless is:
\begin{equation}                                 
m^2 + {1 \over 4} \epsilon^2 g'^2 {v^2 v^2_2 \over v^2 + v^2_2} 
= 0.
\end{equation}
This condition can be satisfied by assigning negative value 
either to $m^2$ or to $v^2_2$.

Assuming for simplicity that the vacuum expectation of the 
second doublet is much
smaller than that of the Higgs doublet, we can write down this
expression for the mixing angle:
\begin{equation}
\sin^2\theta_2=\sin^2\theta(1+2\epsilon_2\cos^2\theta{v_2^2 
\over v^2}),
\end{equation}
where $\theta$ is the Weinberg angle {\em of the Standard Model}.
The electromagnetic interaction is now given by: 
\begin{eqnarray}
{\cal L}^{em}_2 &=& {\cal L}_{2l}^{em} + {\cal L}_{2q}^{em} \\
{\cal L}_{2l}^{em} &=& A_{\mu}[{1 \over 2}(g\sin\theta_2 - 
g'\cos\theta_2)
\bar{\nu}_{L}\gamma^{\mu}{\nu}_L
-{1 \over 2}(g\sin\theta_2 + g'\cos\theta_2) 
\bar{e}_{L}\gamma^{\mu}{e}_L
\nonumber\\
&& - g'\cos\theta_2 \bar{e}_{R}\gamma^{\mu}{e}_R] \\
{\cal L}_{2q}^{em} &=& A_{\mu}[{1 \over 2}
g\sin\theta_2(\bar{u}_{L}\gamma^{\mu}{u}_L -
\bar{d}_{L}\gamma^{\mu}{d}_L) + g'\cos\theta_2({1 \over
6}\bar{u}_{L}\gamma^{\mu}{u}_L +{1 \over
6}\bar{d}_{L}\gamma^{\mu}{d}_L + \nonumber\\
&& {2 \over
3}\bar{u}_{R}\gamma^{\mu}{u}_R - {1 \over
3}\bar{d}_{R}\gamma^{\mu}{d}_R)]
\end{eqnarray}                                 
We see that the charge dequantization and charge non-conservation
effects are controlled by the parameter
$\delta = g \sin \theta_2 -g'\cos \theta_2$.
This parameter measures the deviation of our
theory from the Standard Model (in the latter $g \sin \theta -g'\cos
\theta =0$). Up to the terms of the order of ${v_2^2 \over v^2}$ we
have:
$\delta=e\epsilon_2{v^2_2 \over v^2}$.
In terms of $\delta$ we can conveniently express the dequantized
lepton and quark charges.
The neutrino charge is:
$Q_{\nu}= {1 \over 4} \delta$.
The axial electron charge is equal to:
$Q^5_e= - {1 \over 4} \delta$.
Our normalization is such that the vector electron charge should
 coincide exactly with $-e$, without any corrections:
$Q_e= -e$.
The vector ($Q_u$) and the axial ($Q_u^5$) charges of u-quark are
given by:
$Q_u = {2 \over 3}e + {1 \over 12}\delta$,
$Q_u^5 = {1 \over 4} \delta$.
The charges of d-quark are equal to:
$Q_d =-{1 \over 3}e - {1 \over 6} \delta$,
$Q_d^5 = -{1 \over 4} \delta$.
Consequently, the vector charge of the neutron is:
$Q_n = Q_u + 2Q_d = -{1 \over 4} \delta$.
The vector charge of the proton equals
$Q_p = 2Q_u + Q_d = e$.
Therefore, although the electric charge is dequantized in this model,
nevertheless the following relations between the fermion charges hold
true:
$Q_n + Q_{\nu} =0 ; \;\;\; Q_p + Q_e =0$.

From various experiments testing the validity of electric charge
quantization we can infer the following upper bounds on the parameter
$\delta$.
From the upper bound (\cite{brf,bc}) on the (electron) neutrino 
charge:
 $\delta < 4 \times 10^{-13}$  or $4 \times 10^{-17}$.
From the constraint (\cite{n}) on the neutron electric charge:
$ \delta < 4 \times 10^{-21}$  .
From the tests (\cite{ep}) of the neutrality of atoms:
 $\delta < 4 \times 10^{-18}$ .

5.To summarize, we have analyzed the issues of electric charge
(non)conservation and the photon mass in the context of 
a new class of extended
$SU(2) \times U(1)$
models (the characteristic feature of the class being
the inclusion of a hard mass term for the U(1) gauge
field). We have shown that the massiveness of photon 
necessarily implies  non-conservation (and also dequantization)
of the electric charge
(even in the absence of spontaneous breakdown of the 
electromagnetic symmetry). This situation is in contrast with
the case of the abelian U(1) massive electrodynamics. On the other
hand, we have demonstrated that even in models with 
non-conservation of the electric charge it is possible to
keep the photon mass zero (at least, at the tree level).

The authors are grateful to R.Foot and R.Volkas for stimulating
discussions.

This work was supported in part by the Australian Research Council.


\begin{references}
\bibitem[*]{byline1}e-mail:sasha@tauon.ph.unimelb.edu.au
\bibitem[\dag]{byline2}e-
mail:joshi@bradman.ph.unimelb.edu.au

\bibitem{GN}For a review, see A.S.Goldhaber and M.M.Nieto, Rev. Mod. 
Phys. 
43 (1971) 277.

\bibitem{we1}A.Yu.Ignatiev, V.A.Kuzmin and M.E.Shaposhnikov, Phys.
Lett. 
B84 (1979) 315.
\bibitem{okun}L.B.Okun and Ya.B.Zeldovich, Phys.Lett. B78 (1978) 597;
M.B.Voloshin and L.B.Okun, Pisma v ZhETF, 28 (1978) 156.
\bibitem{we2}A.Yu.Ignatiev, V.A.Kuzmin and M.E.Shaposhnikov, in Proc. 
Int.Conf. Neutrino-79 v.2, p.488. 

\bibitem{s} M.Suzuki, Phys. Rev. {\bf D38} (1988) 1544.
\bibitem{ts} M.M.Tsypin, Yad. Fiz. {\bf 50} (1989) 431.

\bibitem{Qnoncons} K.S.Babu and R.N.Mohapatra, Phys.
Rev.{\bf D42}, 3866 (1990). 
\bibitem{mtt}M.Maruno, E.Takasugi and
M.Tanaka, Progr.Theor.Phys. {\bf 86}, 907 (1991); E.Takasugi
and M.Tanaka, Phys.Rev. {\bf D44}, 3706 (1991).

\bibitem{mn} R.N.Mohapatra and S.Nussinov, Int. J.
Mod. Phys. {\bf A7}, 3817 (1992)

\bibitem{Mel} R.R.Foot,G.C.Joshi, H.Lew and R.R.Volkas,
Mod.Phys.Lett. {\bf A5}, 95 (1990); {\em ibid.\/} {\bf A5},
2721 (1990); X.-G.He, G.C.Joshi, H.Lew and R.R.Volkas,
Phys.Rev.{\bf D43}, R22 (1991); {\em ibid.\/} {\bf D44},
2118 (1991); X.-G.He, G.C.Joshi and B.H.J.McKellar,
Europhysics Lett. {\bf 10}, 709 (1989); K.S.Babu and
R.N.Mohapatra, Phys.Rev.Lett. {\bf 63}, 938 (1989);
Phys.Rev. {\bf D41}, 271 (1990); R.R.Foot, Mod.Phys.Lett.
{\bf A6}, 527 (1991); N.G.Deshpande, Oregon Report OITS-107
(1979) (unpublished); a review: R.R.Foot, H.Lew and R.R.Volkas,
J.Phys.G {\bf 19}, 361 (1993); E.Takasugi and M.Tanaka,
Progr.Theor.Phys. {\bf 87}, 679 (1992);

\bibitem{partdata}Particle Data Group, Review of Particle
Properties, Phys. Rev. {\bf D50} (1994) Part 1.

\bibitem{n} J.Baumann, R.Gahler, J.Kalus, and W.Mampe, Phys.
Rev. {\bf D37}, 3107 (1988); see also R.Gahler, J.Kalus, and
W.Mampe, Phys. Rev. {\bf D25}, 2887 (1982)

\bibitem{ep}M.Marinelli and G.Morpurgo, Phys. Lett. {\bf
137B}, 439 (1984); see also J.C.Zorn, G.E.Chamberlin, and
V.W.Hughes, Phys. Rev. {\bf 129}, 2566 (1963); H.F.Dylla and
J.G.King, Phys. Rev. {\bf A7}, 1224 (1973)

\bibitem{brf} J.Bernstein, M.Ruderman and G.Feinberg,
Phys.Rev. {\bf 132}, 1227 (1963)

\bibitem{bc}G.Barbiellini and G.Cocconi, Nature {\bf 329},
21 (1987)

\bibitem{bv}K.S.Babu and R.R.Volkas, Phys. Rev. {\bf D46},
2764 (1992)

\bibitem{clt}J.M.Cornwall, D.N.Levin and G.Tiktopoulos, 
Phys. Rev. Lett. 32 
(1973) 498; Phys. Rev. {\em D10} (1974) 1145.

\bibitem{delb}R.Delbourgo, S.Twisk and G.Thompson, Int.J.
Mod.Phys.A3, 435 (1988).
\end{references}
\end{document}